\documentclass[aps,prb,twocolumn,groupedaddress,showpacs]{revtex4}

\usepackage{graphicx}
\usepackage{amsmath}

\begin{document}


\title{Instability of vortex array and transitions to turbulent
states in rotating helium~II}


\author{Makoto Tsubota$^1$}
\author{Tsunehiko Araki$^1$}
\author{Carlo F. Barenghi$^2$}

\affiliation{
$^1$Department of Physics,
Osaka City University, Sumiyoshi-Ku, Osaka 558-8585, Japan 
\\
$^2$School of Mathematics and Statistics, University of Newcastle,\\
Newcastle upon Tyne NE1 7RU, UK}


\date{\today}

\begin{abstract}
We consider superfluid helium inside a container which rotates at
constant angular velocity and investigate numerically the stability of the
array of quantized vortices in the presence of an imposed axial
counterflow. This problem was studied experimentally by Swanson
{\it et al.}, who reported evidence of instabilities at
increasing axial flow but were not able to explain their nature.
We find that Kelvin waves on individual vortices
become unstable and grow in amplitude, until the amplitude of the
waves becomes large enough that vortex reconnections take place and
the vortex array is destabilized.  The eventual
nonlinear saturation of the instability consists of a turbulent
tangle of quantized vortices which is strongly polarized.
The computed results compare well with the experiments.
Finally we suggest a theoretical explanation for the second instability which
was observed at higher values of the axial flow.
\end{abstract}


\maketitle


\section{Introduction}

The work described in this article is concerned with the stability of
a superfluid vortex array. It is well known\cite{Donnelly}\cite{BDV2}
that, if helium~II is rotated at constant angular velocity $\Omega$,
an array of superfluid vortex lines is created. The
vortices are aligned along the axis of rotation and form an array
with areal density given by

\begin{equation}
L_{rot}=\frac{2\Omega}{\kappa},
\label{rot0}
\end{equation}

\noindent
where $\kappa=h/m=9.97 \times 10^{-4}{\rm cm^2/sec}$ is the quantum of
circulation, $h$ is Plank's constant and $m$ the mass of one helium atom.
Equation (\ref{rot0}) is valid provided that $\Omega$ exceeds a small
critical value\cite{Fetter}. Rotation frequencies of the order of
$1~\rm Hz$ are easily achieved in a laboratory, and correspond to real
density of the order of $10^3~\rm cm^{-2}$.

It is also well known that a superfluid vortex line becomes unstable in the
presence of normal fluid in the direction parallel to the axis of
the vortex. This instability, hereafter referred to as the Donnelly-
Glaberson (DG) instability, was first observed experimentally
by Cheng {\it et al.} \cite{Cheng} and then
explained by Glaberson {\it et al.} \cite{Glaberson}. Physically, the
DG instability takes the form of Kelvin waves
(helical displacements of the vortex core) which grow exponentially
with time.

In this paper we use an imposed axial flow to trigger the
DG instability and study
the transition from order to disorder in an array of quantized
vortex lines. It is useful to remark here that,
since the growth of Kelvin waves takes place
at the expense of normal fluid's energy, understanding the DG instability is
also relevant\cite{SK} to the balance of energy between normal fluid and
superfluid in helium~II turbulence, a problem which is attracting
current experimental
\cite{Stalp} \cite{Tabeling}\cite{Lancaster1}\cite{Lancaster2}
and theoretical\cite{ABC}\cite{BHS}\cite{Tsubota}\cite{Vinen1}
attention.

The article is organized in the following way. In section 2 we
describe the rotating counterflow configuration, which is
relevant to both theory and experiment.  In section 3 we summarize
experimental results obtained by Swanson, Barenghi and
Donnelly\cite{Swanson}.  They discovered
that the DG mechanism can destabilize the superfluid vortex array
and  revealed the existence of two different superfluid states
at increasing values of the driving axial flow beyond
the DG transition.
Until now, the actual physical nature of these two states
has been a mystery, and it is the aim of our work to shed light into this
problem. In section 4 we set up the formulation of vortex dynamics in
the rotating frame which generalizes the previous approach of Schwarz
\cite{Schwarz} and which we use in our numerical calculations.
Section 5 is devoted to the DG instability. What happens beyond the
DG instability cannot be predicted by linear stability theory
and must be investigated by direct
nonlinear computation, which is done in section 6. In section 7 we
tackle the transition to the second turbulent state
observed by Swanson {\it et al}. Finally, section 8 draws the
conclusions.

\section{Rotating counterflow}

In order to study the stability of the rotating superfluid vortex
array, we consider the configuration which is
schematically shown in Fig. 1.  A channel, which is closed
at one end and open to the helium bath at the other end, is placed
on a table which can be rotated at an assigned angular velocity $\Omega$.
At the closed end of the channel a resistor dissipates a known heat
flux $\dot Q$.

\begin{figure}[tbhp]
\includegraphics[height=0.3\textheight]{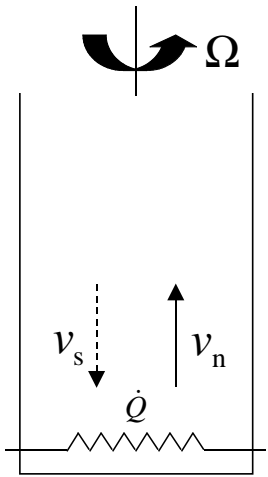}
\caption{Schematic rotating counterflow apparatus.}
\label{eps1}
\end{figure}

First let us consider what happens in the absence of rotation ($\Omega=0$).
Since only the normal fluid carries entropy, then
$\dot Q=\rho T S V_n$, where $T$ is the absolute temperature, $S$ the
specific entropy, $\rho=\rho_s+\rho_n$ the total density of helium~II,
$\rho_s$ the superfluid density and $\rho_n$ the normal fluid density. We call
$V_n$ and $V_s$ respectively the normal fluid and the superfluid velocity
fields in the direction along the channel, averaged over the channel's
cross section.
The total mass flux $\rho_s V_s +\rho_n V_n$ is zero because one end of
the channel is closed.
The resulting counterflow velocity $V_{ns}=V_n-V_s$ which is
induced along the channel is therefore
proportional to the applied heat flux:
\begin{equation}
V_{ns}=\frac{\dot Q}{\rho_s S T}
\label{vnsq}
\end{equation}

It is known from experiments\cite{Tough}\cite{Vinen2} and numerical simulations
\cite{Schwarz} that, if $\dot Q$ (hence $V_{ns}$) exceeds a critical
value, a turbulent tangle of quantized vortex lines is created.
The tangle is homogeneous and isotropic (neglecting a small degree of
anisotropy induced by the direction of the imposed hear current).
The intensity of the turbulence is measured by the vortex line density
(length of vortex line per unit volume) which is experimentally
determined by monitoring the extra attenuation of second sound.
It is found that the vortex line density has the form

\begin{equation}
L_{\rm  flow} =\gamma_H^2 V_{\rm  ns}^2,
\label{heat}
\end{equation}

\noindent
where $\gamma_H$ is a temperature dependent coefficient \cite{Tough}.

Let us consider now the case in which
the heat flux is applied in the rotating frame ($\Omega \ne 0$).
We have now two effects which compete with each other:
rotation, which favours the creation of an ordered array of vortices aligned
along the direction of the axis of rotation, and counterflow,
which favours the creation of a disordered tangle.
Swanson {\it et al.}\cite{Swanson} were the first to address
the problem of whether the vortex array is stable or not at given
values of $\Omega$ and $V_{ns}$, and, if the array is unstable, of whether
the vortex line density $L$ is the sum of Eq.~(\ref{rot0}) and
Eq.~(\ref{heat}) or not. Their experimental results are
described in the next section.

It is important to remark that, in principle,
one can also study the stability of a vortex array in the presence
of a mass flow rather than of a heat current. Similarly, one can study
the effects of rotation upon the turbulence of helium~II created by
towing a grid or rotating a propellers rather than upon counterflow
turbulence.
The reason for which we have chosen to restrict our investigation
to the case of a heat current is twofold: firstly, the experimental
data of Swanson {\it et al}\cite{Swanson}
are available; secondly, at least at small heat
currents\cite{Melotte}, the turbulent superfluid tangle is
homogeneous and almost isotropic and we do not have to worry about
large scale motion and eddies of the normal fluid.

\section{The experiment}

The rotating counterflow apparatus of Swanson {\it et al}
\cite{Swanson} consisted of a $40~\rm cm$ long vertical channel with
$1 \times 1~\rm cm^2$ square cross section.
At the closed end (as shown in Fig. 1) a resistor dissipated
a known heat flux $\dot Q$ and induced relative motion $V_{ns}$ of the
two fluid components. The vortex line density $L$ was measured by pairs
of second sound transducers located along the channel.
The entire apparatus was set up on a rotating cryostat, so that it was
possible to create vortex lines by either rotation or counterflow, or
by any combination of them.
The vortex line density was calculated from a measurement of the attenuation
of second sound resonances and its calibration against the known
density in rotation \cite{BPD}.

The experiment was performed at $T=1.65~\rm K$.  In the presence of
both rotation and counterflow three distinct flow states were observed,
as shown in Fig. 2.
The three states are separated by two critical counterflow velocities
$V_{c1}$ and $V_{c2}$, which are respectively
the boundaries between the primary state and the
secondary state, and between the secondary state and the tertiary state.
The results of the experiment can be summarized as it follows:

\noindent{$\bullet$ \bf Primary state}

In the first region of Figure 2 at the left of $V_{c1}$ the
vortex line density is independent of the small values of $V_{ns}$
involved and agrees with Eq.~(\ref{rot0}). This region clearly
corresponds to an ordered vortex array, and the counterflow current
$V_{ns}$ is not strong enough to destabilize it.

\noindent{{$\bullet$ \bf Transition from primary state to secondary state}

Swanson {\it et al.} \cite{Swanson} noticed that the values of the first
critical velocity $V_{c1}$ are consistent with the DG instability.
This means that at $V_{ns}=V_{DG}$ the axial flow is so strong that
Kelvin waves of infinitesimal amplitude become unstable.

\noindent{{$\bullet$ \bf Secondary state}

Because of the lack of
direct flow visualization in helium~II, the nature of the flow
past the instability ($V_{ns}>V_{c1}$) was not clear to Swanson
{\it et al} \cite{Swanson}.
The only information which they could recover
by the second sound measurement technique was that rotation added
fewer than the expected $2\Omega/\kappa$ vortex lines to those
 made by the counterflow current.

\noindent{{$\bullet$ \bf Transition from secondary state to tertiary state}

The existence of a second critical velocity  $V_{c2}$ was
unexpected. The nature of the transition at $V_{ns}=V_{c2}$ and which kind
of flow exists in the third region ($V_{ns}>V_{c2}$) were
a mystery.

\begin{figure}[tbhp]
\includegraphics[height=0.3\textheight]{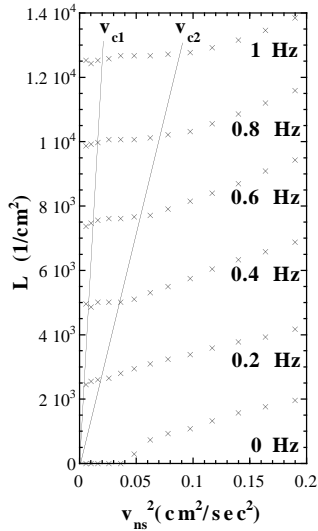}
\caption{Vortex line density $L$ observed by Swanson {\it et al.}
\cite{Swanson}
in the presence of a counterflow heat current $V_{ns}$ at various rotation
rates at $T=1.65$. The solid lines represent fits
to the two observed critical velocities $V_{c1}$ and $V_{c2}$. The
experimental uncertainties are about $1/3$ of the symbol size.}
\label{eps2}
\end{figure}

\section{Vortex dynamics in a rotating frame}

The vortex filament model is very useful to study the motion of superfluid
$^4$He because the vortex core radius $a_0 \sim 10^{-8}$ cm is microscopic,
hence much smaller than any flow scales of interest.
Moreover, unlike what happens in classical fluid dynamics,
the circulation $\kappa = 9.97 \times 10^{-4}$ cm$^2$/sec is
fixed by quantum constraint, which simplifies the model ever further.

Helmholtz's theorem for a perfect fluid states that a vortex
moves with self - induced velocity at each place produced by the shape of
the vortex itself.  Therefore the velocity $\dot{\Vec{s}}_0$ of a vortex
filament at the point $\Vec{s}$ in the absence of mutual friction is
governed by the Biot-Savart law and can be expressed as \cite{Schwarz}:

\begin{equation}
\dot{\Vec{s}}_0 = \frac{\kappa}{4\pi}\Vec{s}' \times \Vec{s}''
\ln \left( \frac{2 (l_{+} l_{-})^{1/2}}{e^{1/4} a_0} \right) +
\frac{\kappa}{4\pi}
\int ' \frac{(\Vec{s}_1 - \Vec{r}) \times d\Vec{s}_1}{|\Vec{s}_1 -
\Vec{r}|^3}.
\label{rot1}
\end{equation}

\noindent
Here the vortex filament is represented by the parametric equation
$\Vec{s}=\Vec{s}(\xi,t)$.
The first term means the localized induction velocity, where
the symbols $l_{+}$ and $l_{-}$ are the lengths of the two line
elements which are  adjacent to a given point after discretization of
the filament, and the prime denotes differentiation of ${\Vec{s}}$
with respect to the arc length $\xi$.
The second term represents the nonlocal field obtained by carrying out the
integral
along the rest of the filament on which $\Vec{s}_1$ refers to a point.

If the temperature is finite, the normal fluid fraction is non-zero and
its effects must be taken into account. The normal fluid induces a mutual
friction force which drags the vortex core of a superfluid vortex filament
for which the velocity of point $\Vec{s}$ is given by

\begin{equation}
\dot{\Vec{s}}=\dot{\Vec{s}}_0 + \alpha \Vec{s}' \times (\Vec{v}_{\rm ns} -
\dot{\Vec{s}}_0)- \alpha '
\Vec{s}' \times [\Vec{s}' \times (\Vec{v}_{\rm ns} - \dot{\Vec{s}}_0)],
\label{rot2}
\end{equation}

\noindent
where $\alpha$ and $\alpha'$ are known temperature-dependent friction
coefficients and $\dot{\Vec{s}}_0$ is calculated from Eq. (\ref{rot1}).
More details of the numerical method and how it is implemented are
described in Ref.\cite{TAN}.

In order to make progress in our problem, we need to generalize this vortex
dynamics approach to a rotating frame \cite{Yoneda}.
The natural way to perform the calculation in a rotating frame would
require to consider a cylindrical container. We do not
follow this approach for two reasons.
Firstly, our formulation is implemented using the
full Biot - Savart law, not the localized-induction approximation
often used in the literature.
This would require to place image
vortices beyond the solid boundary to impose the condition of no flow
across it.  This is easily done in cartesian (cubic) geometry,
but it is difficult to do in  cylindrical geometry,
Secondly, the original experiment by Swanson {\it et al.} \cite{Swanson} was
carried out in a rotating channel with a square cross section.

In a rotating vessel the equation of motion of vortices is modified by two
effects.
The first is the force acting upon the vortex due to the rotation.
According to the Helmholtz's theorem, the generalized force acting upon the
vortex
is balanced by the Magnus force:
\begin{equation}
\rho_{\rm s}\kappa (\Vec{s}' \times \dot{\Vec{s}}_0)=\frac{\delta F'}{\delta
\Vec{s}}, \label{rot3}
\end{equation}
where $F'=F-\Vec{\Omega}\cdot \Vec{M}$ is the free energy of a system in a
frame rotating
around a fixed axis with the angular velocity $\Vec{\Omega}$ and the angular
momentum $\Vec{M}$.
Taking the vector product of Eq.(\ref{rot3}) with $\Vec{s}'$, we obtain the
velocity
$\dot{\Vec{s}}_0$.
The first term $F$ due to the kinetic energy of vortices gives that
Biot - Savart law, and the second
term $\Vec{\Omega}\cdot \Vec{M}$ leads to the velocity $\dot{\Vec{s}}_{\rm
rot}$ of the vortex caused by the rotation:

\begin{eqnarray}
\dot{\Vec{s}}_{\rm rot}  = \frac{1}{4\pi} & \int & \left\{ 3\frac{\Vec{s}'
\times
\Vec{R}}{|\Vec{R}|^5} [(\Vec{\Omega}\cdot
\Vec{s}')(\Vec{r}\cdot
 \Vec{R})-(\Vec{\Omega}\cdot \Vec{R})(\Vec{r}\cdot
\Vec{s}')] \right. \nonumber \\
& & {}  + \frac{\Vec{s}' \times
\Vec{\Omega}}{|\Vec{R}|^5}[|\Vec{R}|^2
(\Vec{r}\cdot \Vec{s}') - 3 (\Vec{r}\cdot
\Vec{R}) ~ (\Vec{R}\cdot \Vec{s}')] \nonumber \\
& & {} - \frac{\Vec{s}' \times \Vec{r}}{|\Vec{R}|^5}
 [|\Vec{R}|^2 (\Vec{\Omega} \cdot \Vec{s}') - 3 (\Vec{\Omega}\cdot
\Vec{R}) ~ (\Vec{R} \cdot \Vec{s}')] \nonumber \\
& & {} \left. - \frac{\Vec{\Omega} \times \Vec{r}}{|\Vec{R}|^3} +
\frac{\Vec{s}'
\cdot( \Vec{\Omega} \times \Vec{r})}{|\Vec{R}|^3} \Vec{s}' \right\}
d\Vec{r} \label{rot4}
\end{eqnarray}
with $\Vec{R}=\Vec{r}-\Vec{s}$.
The second effect is the superflow induced by the rotating vessel.
For a perfect fluid we know the analytical solution of the velocity inside a
cube
of size $D$ rotating with the angular velocity $\Vec{\Omega}=\Omega
\hat{\Vec{z}}$ \cite{Thomson}:
\begin{eqnarray}
\Vec{v}_{{\rm cub},x} &=& \frac{8\Omega}{\pi^2}
\sum_{n=0}^{\infty}\frac{(-1)^n}{(2n+1)^2}\frac{D}{2}
{\rm sech}\frac{(2n+1)\pi}{2} \nonumber \\
& & {} \times \left[ {\rm sinh} Y {\rm
cos} X
 - {\rm cosh} X {\rm sin} Y \right]
\label{rot5} \\
 \Vec{v}_{{\rm cub},y} &=& \frac{8\Omega}{\pi^2}
\sum_{n=0}^{\infty}\frac{(-1)^n}{(2n+1)^2}\frac{D}{2}
 {\rm sech}\frac{(2n+1)\pi}{2} \nonumber \\
& & {} \times \left[ {\rm cosh} Y {\rm
sin} X
 - {\rm sinh} X {\rm cos} Y \right]
\label{rot6}
\end{eqnarray}
with $X=(2n+1) \pi x/D$ and $Y=(2n+1) \pi Y/D$.
In a rotating frame these terms are added to the velocity $\dot{\Vec{s}}_0$
without the mutual friction,
so Eq. (\ref{rot1}) is replaced by
\begin{eqnarray}
\dot{\Vec{s}}_0 &=& \frac{\kappa}{4\pi}\Vec{s}' \times \Vec{s}'' \ln \left(
\frac{2 (l_{+} l_{-})^{1/2}}
{e^{1/4} a_0} \right) + \frac{\kappa}{4\pi} \int ' \frac{(\Vec{s}_1 -
\Vec{r}) \times d\Vec{s}_1}{|\Vec{s}_1 - \Vec{r}|^3} \nonumber \\
& & {} + \dot{\Vec{s}}_{\rm rot} + \Vec{v}_{\rm cub} .
\label{rot7}
\end{eqnarray}

Some important quantities useful for characterizing the rotating tangle will
be introduced.
The vortex line density is
\begin{equation}
L=\frac{1}{\Lambda}\int d\xi,
\end{equation}
where the integral is made along all vortices in the sample volume $\Lambda$.
The polarization of the tangle may be measured by the quantity
\begin{equation}
<s'_z>=\frac{1}{\Lambda L}\int d\xi \Vec{s}'(\xi)\cdot \hat{\Vec{z}},
\label{polarization}
\end{equation}
as a function of time.

The actual numerical technique used to perform the simulation has been
alreay described\cite{TAN}. Here it is enough to say that
a vortex filament is represented by a single string of points at a distance
$\Delta \xi$ apart.  When two vortices approach within $\Delta \xi$, they
are assumed to reconnect\cite{Koplik}.
The computational sample is taken to be a cube of size $D=1.0$ cm.
We adopt periodic boundary conditions along the rotating axis and rigid
boundary conditions
at the side walls.
All calculations are made under the fully Biot-Savart law, placing image
vortices beyond the
solid boundaries.
The space resolution is $\Delta \xi =1.83 \times 10^{-2}$ cm and the
time resolution is
$\Delta t=4.0 \times 10^{-3}$ sec.
All results presented in this paper refer to calculations made in
the rotating frame.
To make comparison with the experiment \cite{Swanson}, we use $\alpha = 0.1$
and $\alpha'=0$ at the temperature $T=1.6$K.
The uniform counterflow ${\Vec{v}_{\rm ns}}$ is applied along the $z$ axis.

\section{The DG instability}

Swanson {\it et al.} \cite{Swanson} found that the first critical velocity
$V_{c1}$ was proportional to $\Omega^{1/2}$; this functional dependence and
the actual numerical values were consistent with interpreting the transition
at $V_{ns}=V_{c1}$ as the DG instability of Kelvin waves.
Glaberson {\it et al.} \cite{Glaberson} considered an array
of quantized vortices (which they modelled as a continuum) inside
a container rotating at an angular velocity $\Omega$.
They found that, in the absence of friction, the dispersion relation
of a Kelvin wave of wavenumber $k$ is

\begin{equation}
\omega=2\Omega +\nu k^2,
\label{omega}
\end{equation}

\noindent
where $\omega$ is the angular frequency of the Kelvin wave,
$\nu$ is given by

\begin{equation}
\nu=\frac{\kappa}{4 \pi} \ln{(\frac{b}{a_0})},
\label{nu}
\end{equation}

\noindent
where $b\approx L^{-1/2}$ is the average distance
between vortices.

Glaberson {\it et al.} \cite{Glaberson} showed that the
dispersion law~(\ref{omega}) has a critical velocity

\begin{equation}
V_{DG}=\frac{\omega}{k}_{min}=2(2\Omega \nu)^{1/2}
\label{vcrit}
\end{equation}

\noindent
at the critical wavenumber

\begin{equation}
k_{DG}=\sqrt{\frac{2\Omega}{\nu}}.
\label{kcrit}
\end{equation}

If the axial flow $V_{\rm ns}$ exceeds $V_{DG}$ for some value of $k$, then
Kelvin waves with that wavenumber $k$ (which are always present at
very small amplitude due to thermal excitations and mechanical vibrations)
become unstable and
grow exponentially in time. Physically, the phase velocity of the
mode $k$ is equal to the axial flow, so energy is fed into the Kelvin wave
by the normal flow.

Figure 3 illustrates the DG instability. The
computations were performed in a periodic box of size $1~\rm cm$
in a reference frame rotating with angular velocity
$\Omega=9.97 \times 10^{-3}~\rm rad/sec$, for which
$V_{DG}=0.010~\rm cm/sec$.
Figure 3a confirms that when $V_{ns}=0.008~\rm cm/sec<V_{DG}$ the
vortex lines remain stable. Figure 3b shows that, at
$V_{ns}=0.015~\rm cm/sec>V_{DG}$, Kelvin waves become unstable and grow,
as predicted.
Figures 3c,d,e and f show that Kelvin waves of larger wavenumber become
unstable
at higher counterflow velocity.

\begin{figure}[tbhp]
\begin{minipage}{1.0\linewidth}
\includegraphics[height=0.14\textheight]{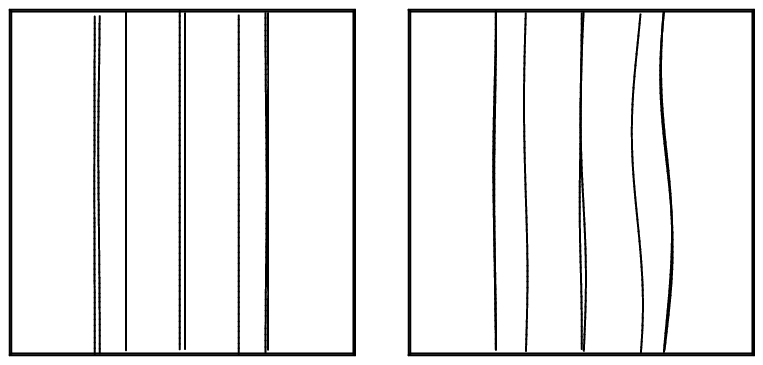} \\
(a) \hspace{2.5cm} (b)
\end{minipage}
\begin{minipage}{1.0\linewidth}
\includegraphics[height=0.14\textheight]{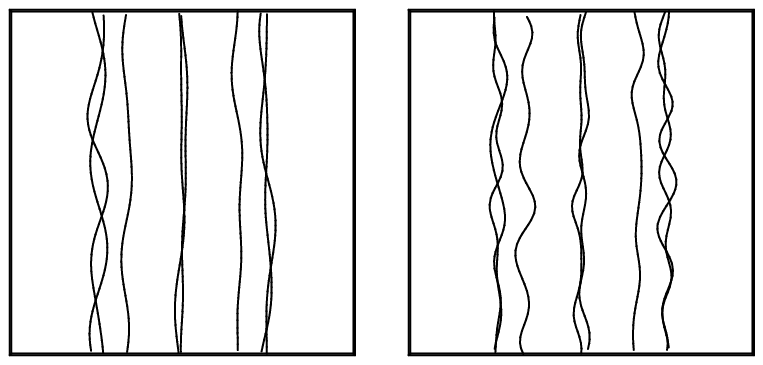} \\
(c) \hspace{2.5cm} (d)
\end{minipage}
\begin{minipage}{1.0\linewidth}
\includegraphics[height=0.14\textheight]{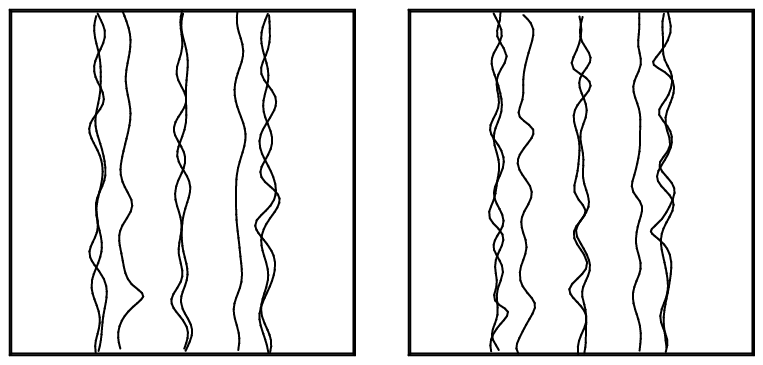} \\
(e) \hspace{2.5cm} (f)
\end{minipage}
\caption{Numerical simulations of the Donnelly-Glaberson
instability at $\Omega=9.97 \times 10^{-3} {\rm rad/sec}$,
$T=1.6{\rm K}$.
Snapshots of vortex configurations at the following
counterflow velocities $V_{ns}$:
(a): $V_{ns}=0.008{\rm cm/sec}$;
(b): $V_{ns}=0.015{\rm cm/sec}$;
(c): $V_{ns}=0.03{\rm cm/sec}$;
(d): $V_{ns}=0.05{\rm cm/sec}$;
(e): $V_{ns}=0.06{\rm cm/sec}$;
(f): $V_{ns}=0.08{\rm cm/sec}$.}
\label{eps3}
\end{figure}

Linear stability theory\cite{Drazin} can only predict two quantities:
the first is the critical
value of the driving parameter ($V_{DG}$ in our case) at which
a given state (the vortex array in our case) becomes unstable
because infinitesimal perturbations grow rather than decay;  the second is the
exponential growth or decay rate of these perturbations for a given
value of the driving parameter. Therefore the linear
stability theory of Glaberson\cite{Glaberson}
cannot answer the question of what is the new solution which
grows beyond the DG instability: to determine this new solution (what we
call the secondary state in section 3) we must
solve the governing nonlinear equations of motion, which is what we
do in the next section.

\section{Rotating turbulence}

Because of the computational cost of the Biot-Savart law, it is
not practically possible to compute vortex tangle with densities
which are as high ($L={\cal O}(10^4){\rm cm^{-2}}$)
as those achieved in the experiment.
Nevertheless, numerical simulations performed at smaller, computationally
realistic values of $L$ are sufficient to shed light into
the physical processes involved. Some results
which we describe have been already presented in preliminary
form\cite{preliminary}; together with more recent computer simulations,
the picture which emerges and which we present here
gives a good understanding of the experimental findings
of Swanson {\it et al}\cite{Swanson}, at least  as far as the transition to the
secondary flow and the secondary flow itself are concerned.

The time sequence contained in  Fig. 4 illustrates the evolution
of a  vortex array at a relatively small angular velocity
$\Omega=9.97 \times 10^{-3}{\rm rad/sec}$, in the presence of
the counterflow $V_{ns}=0.08{\rm cm/sec}$. Figure 4a shows the initial $N=8$
parallel vortex lines at $t=0$. The vortices have been seeded with
small random perturbations to make the simulation more realistic.
The absence of these perturbations would make the phase of the Kelvin waves
synchronize on all vortices to delay reconnections.
As the evolution proceeds, perturbations with  high wavenumbers
are damped by the mutual friction, whereas perturbations which are linearly
DG-unstable
grow exponentially, hence Kelvin waves become visible (Fig. 4b).
When the amplitude of the Kelvin waves becomes of the order of the average
vortex separation, reconnections take place (Fig. 4c).
The resulting vortex loops disturb the initial vortex array, leading to an
apparently random vortex tangle (Fig. 4d).
After the initial exponential growth (which is predicted by the theory of the
DG instability), nonlinear effects (vortex interactions and vortex
reconnections) become important and nonlinear saturation takes
place.

\begin{figure}[tbhp]
\begin{minipage}{1.0\linewidth}
\includegraphics[height=0.12\textheight]{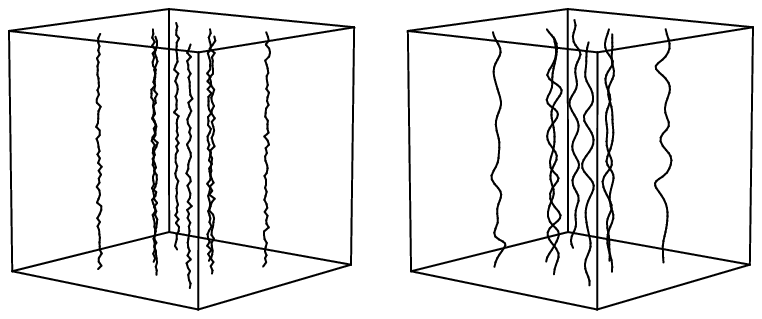} \\
(a) \hspace{2.5cm} (b)
\end{minipage}
\begin{minipage}{1.0\linewidth}
\includegraphics[height=0.12\textheight]{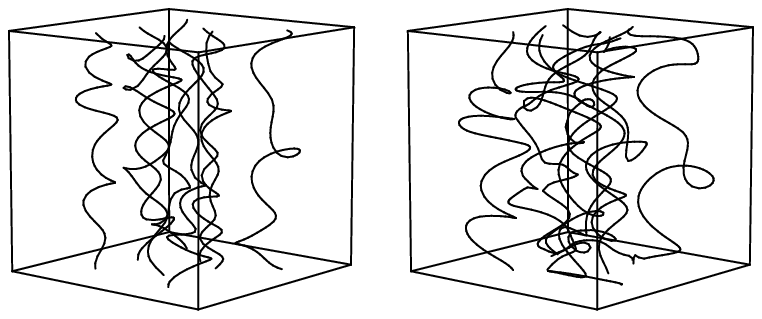} \\
(c) \hspace{2.5cm} (d)
\end{minipage}
\begin{minipage}{1.0\linewidth}
\includegraphics[height=0.12\textheight]{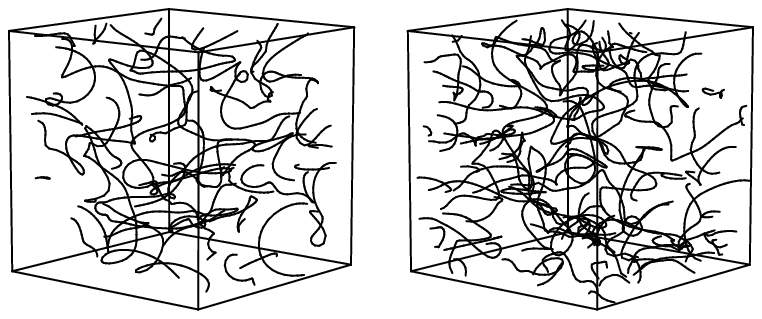} \\
(e) \hspace{2.5cm} (f)
\end{minipage}
\caption{Numerical simulation of rotating turbulence at
$T=1.6{\rm K}$, $\Omega=9.97 \times 10^{-3}{\rm rad/sec}$
and $V_{ns}=0.08{\rm cm/sec}$. Computed vortex tangle at
the following times:
(a): t=0{rm sec};
(b): t=16{\rm sec};
(c): t=28{\rm sec};
(d): t=36{\rm sec};
(e): t=80{\rm sec};
(f): t=600{\rm sec}.}
\label{eps4}
\end{figure}

Figure 5 shows a similar time sequence at the same counterflow velocity
$V_{ns}=0.08{\rm cm/sec}$
but at higher rotation rate $\Omega=4.98 \times 10^{-2}{\rm rad/sec}$.
In this case we have $N=33$ initial parallel vortices (Fig. 5a).
At $t=12{\rm sec}$ (Fig. 5b) it is still $N=33$.
Then the amplitude of the Kelvin waves becomes so large that lots of
reconnections
take place and $N$ increases; for example, we have $N=83$ at $t=160{\rm
sec}$ (Fig. 5f).

\begin{figure}[tbhp]
\begin{minipage}{1.0\linewidth}
\includegraphics[height=0.12\textheight]{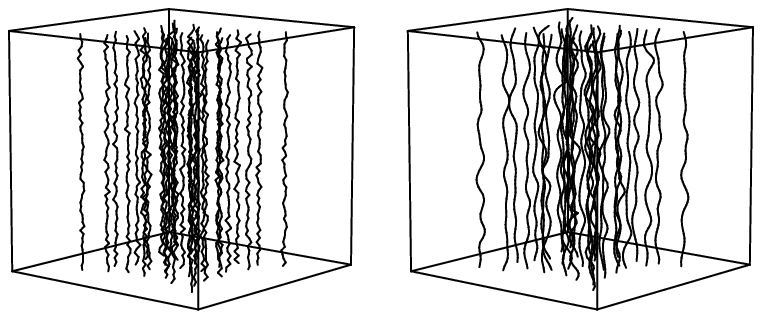} \\
(a) \hspace{2.5cm} (b)
\end{minipage}
\begin{minipage}{1.0\linewidth}
\includegraphics[height=0.12\textheight]{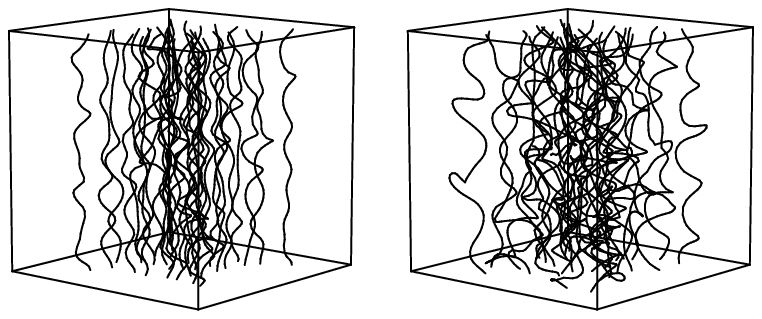} \\
(c) \hspace{2.5cm} (d)
\end{minipage}
\begin{minipage}{1.0\linewidth}
\includegraphics[height=0.12\textheight]{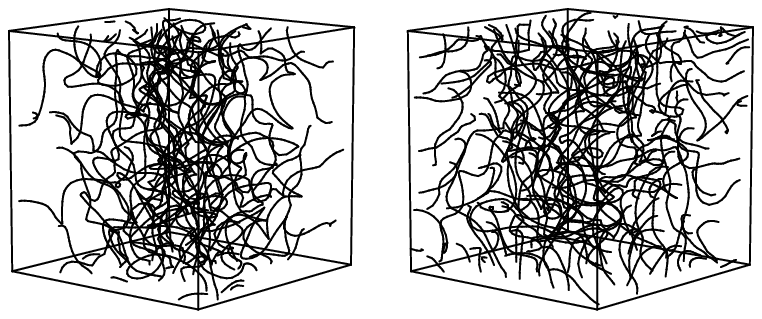} \\
(e) \hspace{2.5cm} (f)
\end{minipage}
\caption{Numerical simulation of rotating turbulence at
$T=1.6{\rm K}$, $\Omega=4.98 \times 10^{-2}{\rm rad/sec}$
and $V_{ns}=0.08{\rm cm/sec}$. Computed vortex tangle at
the following times:
(a): t=0{rm sec};
(b): t=12{\rm sec};
(c): t=20{\rm sec};
(d): t=28{\rm sec};
(e): t=40{\rm sec};
(f): t=160{\rm sec}.}
\label{eps5}
\end{figure}

It is instructive to compare these results with ordinary counterflow
in the absence of rotation. Figure 6 shows a vortex tangle obtained
for $\Omega=0$ and $V_{ns}=0.08{\rm cm/sec}$.
The dynamics starts from $N=6$ vortex rings.
It has been known since the early work of Schwarz \cite{Schwarz}
that the resulting tangle does not depend on the initial condition.
In this particular simulation the vortices develop to a turbulent tangle.

\begin{figure}[tbhp]
\begin{minipage}{1.0\linewidth}
\includegraphics[height=0.12\textheight]{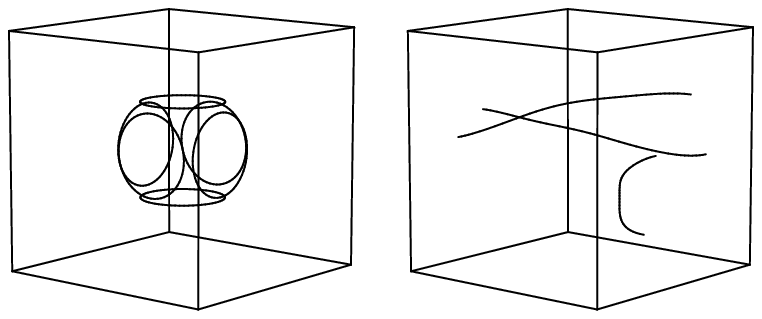} \\
(a) \hspace{2.5cm} (b)
\end{minipage}
\begin{minipage}{1.0\linewidth}
\includegraphics[height=0.12\textheight]{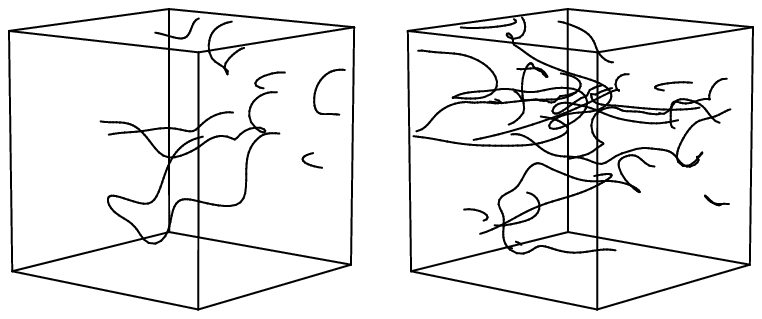} \\
(c) \hspace{2.5cm} (d)
\end{minipage}
\begin{minipage}{1.0\linewidth}
\includegraphics[height=0.12\textheight]{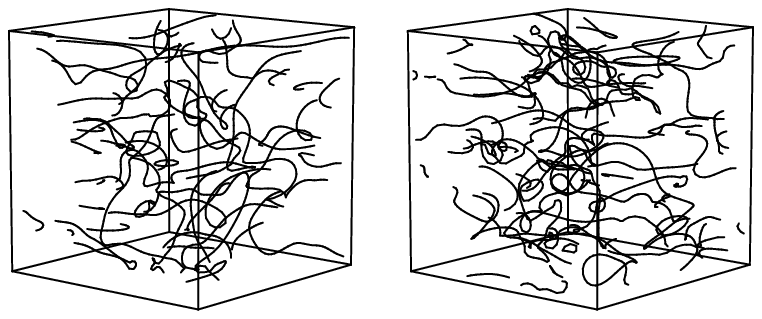} \\
(e) \hspace{2.5cm} (f)
\end{minipage}
\caption{Numerical simulation of counterflow turbulence at
$T=1.6{\rm K}$ in the absence of rotation ($\Omega=0$) for
$V_{ns}=0.08{\rm cm/sec}$. Computed vortex tangle at
the following times:
(a): t=0{rm sec};
(b): t=120{\rm sec};
(c): t=360{\rm sec};
(d): t=520{\rm sec};
(e): t=680{\rm sec};
(f): t=1160{\rm sec}.}
\label{eps6}
\end{figure}

Figure 7 shows that in all three cases (small rotation, large rotation,
no rotation) the vortices, after an initial transient, saturate to a
statistically steady, turbulent state, which is
characterized by a certain average value of
$L$. In the case of $\Omega \neq 0$ (Fig. 6a and b),
it is apparent that the initial growth is exponential, which confirms the
occurrence of a linear instability.

\begin{figure}[tbhp]
\includegraphics[height=0.4\textheight]{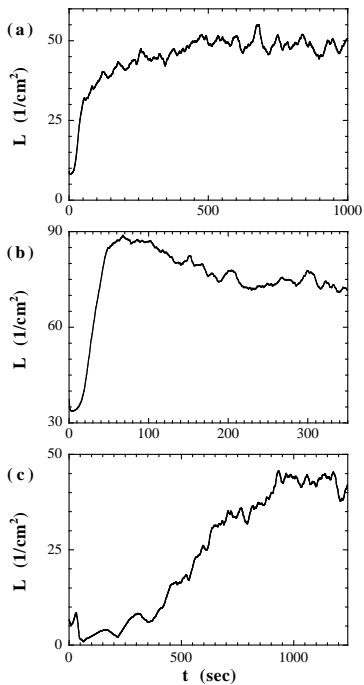} 
\caption{Vortex line density $L$ vs time $t$
at $T=1.6{\rm K}$ and $V_{ns}=0.08{\rm cm/sec}$ for:
(a): $\Omega=9.97\times 10^{-3}{\rm rad/sec}$;
(b): $\Omega=4.98\times 10^{-2}{\rm rad/sec}$;
(c): $\Omega=0{\rm rad/sec}$.}
\label{eps7}
\end{figure}
Looking carefully at the saturated tangle at higher rotation (Fig. 5f)
we notice that there are more loops
oriented vertically than horizontally. The effect is not visible
at lower rotation (Fig. 4f) and at zero rotation (Fig. 6f).
The degree of polarization of the tangle is represented by  $\ <s'_z>$
of Eq.(\ref{polarization}).
This quantity captures the difference between a vortex array
(for which $<s'_z>=1$ because all lines are aligned in the $+z$ direction)
and a random vortex tangle (for which $<s'_z>=0$ because
there is an equal amount of vorticity in the $+z$ and $-z$ directions).
Figure 8 shows how $<s'_z>$ changes with time in the three cases
(small rotation, large rotation, no rotation) considered. The quantities
of interest are the values of $<s'_z>$ at large times in the saturated
regimes.
In the absence of rotation (Fig. 8c)
$<s'_z>$ is negligible but not exactly zero ($<s'_z>\approx 0.01$),
certainly because the driving counterflow is along the $z$ direction.
This small anisotropy of the counterflow
tangle has been already reported in the literature \cite{anisotropy}.
At small rotation (Fig. 8a) there is a small but finite polarization
$<s'_z>\approx 0.15$, whereas at higher rotation (Fig. 8b) the polarization
is significant ($<s'_z>\approx 0.45$) - in fact it is even visible
with the naked eye (Fig. 5f).

\begin{figure}[tbhp]
\includegraphics[height=0.4\textheight]{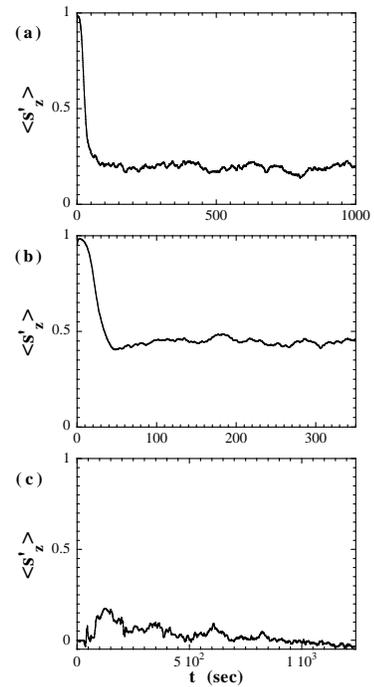} 
\caption{Tangle's polarization $<s'_z>$ vs time $t$
at $T=1.6{\rm K}$ and $V_{ns}=0.08{\rm cm/sec}$ for:
(a): $\Omega=9.97\times 10^{-3}{\rm rad/sec}$;
(b): $\Omega=4.98\times 10^{-2}{\rm rad/sec}$.
(c): $\Omega=0{\rm rad/sec}$.}
\label{eps8}
\end{figure}

Figure 9 shows the calculated dependence of the vortex line density $L$
on the counterflow velocity $V_{ns}$ at different rotation rates $\Omega$.
The figure shows a dependence of $L$ on $V_{ns}$ which is similar to
what appears in the Fig. 1 of the paper by Swanson {\it et al}
\cite{Swanson}.
The only difference is that the scale of the axes in the paper by
Swanson {\it al.} is bigger - in this particular figure they report
vortex line densities as high as $L \approx 2500\rm cm^{-2}$, whereas
our calculations are limited to $L\approx 80\rm cm^{-2}$.
Despite the lack of overlap between the experimental and numerical ranges,
there is clear qualitative similarity between the figures.
It is apparent that the critical velocity beyond which $L$ increases
with $V_{ns}$ is much reduced by the presence of rotation,
which is consistent with the observations.

Figure 10 shows the calculated polarization $<s'_z>$ as a function of
counterflow velocity $V_{ns}$ at different rotation rates $\Omega$.
It is apparent that the polarization decreases with the counterflow
velocity and increases with the rotation, which shows the competition
between order induced by rotation and disorder induced by flow.

\begin{figure}[tbhp]
\includegraphics[height=0.2\textheight]{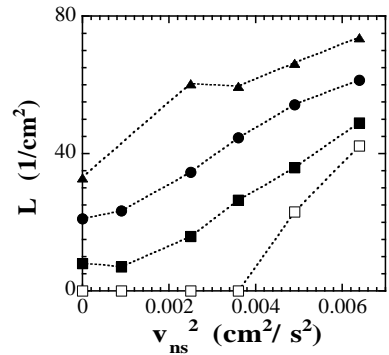} 
\caption{Vortex line density $L$ vs $V_{ns}^2$
at $T=1.6{\rm K}$ for $\Omega=0$ (white square), $\Omega=9.97 \times
10^{-3}{\rm rad/sec}$ (black square),
  $\Omega=2.99 \times 10^{-2}{\rm rad/sec}$ (circle) and
$\Omega=4.98\times 10^{-2}{\rm rad/sec}$ (triangle).
The dotted lines are guides to the eye.}
\label{rot9}
\end{figure}

\begin{figure}[tbhp]
\includegraphics[height=0.2\textheight]{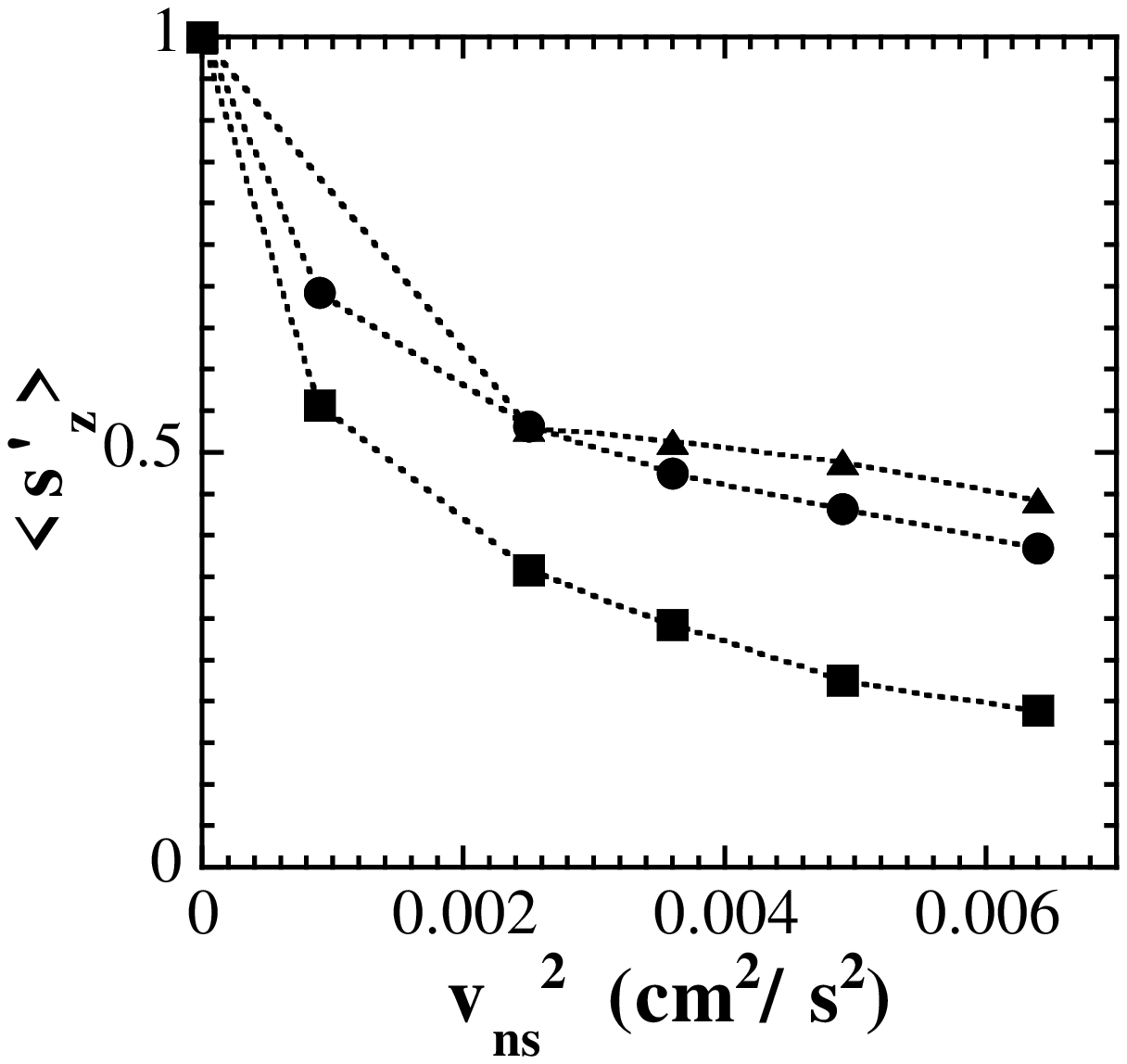} 
\caption{Tangle's polarization $<s'_z>$ vs $V_{ns}^2$
 at $T=1.6{\rm K}$ for $\Omega=9.97 \times 10^{-3}{\rm rad/sec}$ (square),
  $\Omega=2.99 \times 10^{-2}{\rm rad/sec}$ (circle) and
$\Omega=4.98\times 10^{-2}{\rm rad/sec}$ (triangle).The dotted lines
are guides to the eye.}
\label{rot10}
\end{figure}

We conclude that the
nonlinear saturation which takes place beyond the DG instability and
which was observed by Swanson {\it et al.}\cite{Swanson} is
a state of "polarized" turbulence.

\section{The second critical velocity}
In this section we propose a qualitative theory for the
second critical velocity which was observed, but not explained,
by Swanson {\it et al.} \cite{Swanson}. Unfortunately,
this region of parameter
space cannot be explored directly using numerical methods, due to
the larger vortex line densities.

We consider the following idealized model of the region $V_{ns}>V_{c2}$.
We represent the polarized
tangle as the combination of a vortex array and a number of vortex loops
which are randomly oriented, so that the combined system is in balance
and has the necessary amount of length and polarization.
Let $\tau_1$ be the characteristic
timescale of the growing Kelvin waves which are induced on the vortex
array by the DG instability.
The typical lifetime $\tau_2$ of the vortex loops will be
determined by the friction with the normal fluid and by the relative
orientation with respect to the counterflow.
If $\tau_2>\tau_1$ then the vortex loops will not have enough time to shrink
significantly before more loops are introduced by vortex reconnections
induced by growing Kelvin waves. The combined system
consisting of the vortex array and the vortex loops will not be balanced
any longer because randomness is introduced by vortex reconnections at a
rate which is faster than the rate at which loops are destroyed by friction.
In conclusion, we expect that the vortex tangle will be turbulent
and unpolarized.  According to this scenario, the order of magnitude of the
critical velocity $V_{c2}$ is given by the condition
\begin{equation}
\tau_1=\tau_2.
\label{vc2}
\end{equation}

First we estimate $\tau_1$ using a simple model.
For the sake of simplicity we assume an isolated vortex line of helical shape
$\Vec{s}=(\epsilon \cos{\phi};\epsilon \sin{\phi};z)$
where $\phi=kz-\omega t$ and $\epsilon<<1$, hence
$z\approx \xi$  is the arc-length.
The tangent unit vector is  $\Vec{s}'=d{\Vec{s}}/d\xi \approx d{\Vec{s}}/dz=
(-k\epsilon \sin{\phi};k\epsilon \cos{\phi};1)$
and
${\Vec{s}}''=(-k^2\epsilon\cos{\phi};-k^2\epsilon\sin{\phi};0)$.
Using the localized-induction approximation,
the self-induced velocity of the line  at the point ${\bf s}$ is given
by

\begin{equation}
\Vec{v}_i=\nu'\Vec{s}'\times\Vec{s}'',
\label{LIA}
\end{equation}

\noindent
where
$\nu'=\kappa {\cal L}_1/(4\pi)$ and the slowly varying term
${\cal L}_1=\ln{[1/(ka_0)]}$ is assumed constant. Neglecting higher order
terms in $\epsilon$ we have
$\Vec{v}_i= \nu' k^2 \epsilon (\sin{\phi};-\cos{\phi};0)$.

In the absence of friction the equation of motion is simply
$d{\Vec{s}}/dt={\Vec{v}}_i$, hence,
assuming that $\epsilon$ is constant, we find
that the Kelvin wave oscillates with angular frequency
$\omega=\nu' k^2$. This result differs from Glaberson's Eq.~(\ref{omega})
because we
perturbed a single vortex line rather than a continuum of vorticity $2\Omega$
described by the Hall-Vinen equations in the rotating frame
(hence the presence of a different
upper cutoff which makes $\nu'$ different from $\nu$
and the contribution $2\Omega$ to $\omega$).

In the presence of friction, neglecting the small mutual friction coefficient
$\alpha'$ for simplicity, the equation of motion is
$d{\Vec{s}}/dt=\Vec{v}_i+\alpha \Vec{s}' \times (\Vec{v}_{\rm ns}-\Vec{v}_i)$.
Assuming $\Vec{v}_{\rm ns}=(0;0;V_{ns})$ and $\epsilon=\epsilon(t)$, we find
that $d\epsilon/dt=\alpha (kV_{ns}-\nu'k^2)\epsilon$ hence
$\epsilon(t)=\epsilon(0)exp(\sigma t)$ where the growth rate is
$\sigma=\alpha(k V_{ns}-\nu'k^2)$.
Given $V_{ns}$, the largest growth rate occurs for $k=V_{ns}/(2\nu')$
and takes $\sigma=\alpha V_{ns}^2/(4\nu')$, for which we
conclude that

\begin{equation}
\tau_1=\frac{1}{\sigma}=\frac{4\nu'}{\alpha V_{ns}^2}.
\label{tau1}
\end{equation}

To estimate $\tau_2$ we approximate the vortex loops as vortex rings of
radius approximately determined by the average vortex spacing
$\delta\approx L^{-1/2}$.
The characteristic lifetime of a ring of radius $R$ in the
presence of friction is \cite{BDV1}

\begin{equation}
\tau_2=\frac{2 \rho_s \pi R^2}{\gamma {\cal L}_2},
\label{tau2}
\end{equation}

\noindent
where ${\cal L}_2=\ln{[(8R/a_0)-1/2]}$ and $\gamma$ is a known
friction coefficient \cite{BDV1}. Setting $2R= \delta=L^{-1/2}$,
we conclude that the polarized tangle is unstable if

\begin{equation}
L<C_2 V_{ns}^2,
\label{result}
\end{equation}

\noindent
with

\begin{equation}
C_2=\frac{\alpha \pi^2 \rho_s}{2 \gamma \Gamma {\cal L}_1{\cal L}_2}.
\label{c}
\end{equation}

Equation ~(\ref{result}) has the same dependence of $L$ on $V_{ns}$ as that
observed experimentally.  At  $T=1.65 {\rm K}$ we have \cite{BDV1}
$\rho_s=0.1168 {\rm g/cm^3}$, $\gamma=1.3 \times 10^{-5} {\rm g/cm~sec}$,
$\alpha=0.11$. Since $a_0\approx 10^{-8} {\rm cm}$ and the slowly
varying logarithm terms are approximately
${\cal L}_1\approx{\cal L}_2\approx 10$, we conclude that
$C_2\approx 5\times 10^4 {\rm cm^{-4}sec^2}$,
which is of the same order of magnitude of the value
$C_2=16 \times 10^4 {\rm cm^{-4}sec^2}$ found by
Swanson {\it et al}\cite{Swanson}.

\section{Conclusions}

In conclusion, we have studied the stability of a superfluid vortex
array in the presence of an applied counterflow, giving answers to some
questions which were first asked by the pioneering experiment of Swanson
{\it et al}\cite{Swanson}.  After investigating the
DG instability, we have determined the existence
of a new state of superfluid turbulence (polarized turbulence)
which is characterized by two statistically steady state properties,
the vortex line density and the degree of polarization. Although
our computed range of vortex line densities does not overlap with the
much higher values obtained in the experiment, we find the same
qualitative dependence of vortex line density versus counterflow
velocity at different rotations. We have also made some qualitative
progress to understand what happens beyond $V_{c2}$.

Further work with more computing power will hopefully investigate
other aspects of the problem, particularly what happens at high
counterflow velocities and line densities.
We also hope that our work will stimulate more
experiments on this problem.

It is somewhat surprising that so little is known about the destabilization
of a rotational vortex array by an imposed counterflow.
For example, it should be possible to observe the
polarization of turbulence by using simultaneous measurements of
second sound attenuation along and across the rotation axis.

Finally, our work should be of interest to other investigations of
vortex arrays and how they can be destabilized in other systems,
ranging from superfluid $^3$He \cite{Helsinki} to atomic Bose-Einstein
condensates \cite{TKU}. It is also worth noticing
that this study has revealed the crossover of the dimensionality of
vortex systems.
If one considers the three regimes in Fig. 2 one notices
that, at a fixed value of $V_{ns}$,
increasing the rotation rate makes the vortices polarized,
changing the dynamics from three-dimensional to two-dimensional.
This reduction of the dimensionality of turbulence has been observed in
classical fluid
mechanics \cite{Hossain}.


\centerline{\bf ACKNOWLEDGEMENTS}
\medskip

The authors thank W.F.Vinen for useful discussions.
C.F.B. is grateful to the Royal Society for financially supporting this
project.
M.T. is grateful to Japan Society for the Promotion of Science for
financially supporting this Japan-UK Scientific Cooperative Program(Joint
Research Project).

\end{document}